\documentclass[aps,prl,final,twocolumn,letterpaper]{revtex4-1}

\usepackage{graphicx}   
\usepackage{import}                         
\usepackage{epstopdf}
\usepackage{amsmath} 
\usepackage{float} 
\usepackage{bm}
\usepackage{amssymb}
\usepackage{quotes}
\usepackage{indentfirst}
\usepackage{color}
\usepackage{transparent}
\usepackage{dcolumn}
\usepackage{braket}
\usepackage{multirow}
\usepackage{cancel} 
\usepackage{mdframed}
\usepackage{color}
\usepackage{bm}
\usepackage{dsfont}
\usepackage{slashed}
\usepackage{soul, color} 
\soulregister\ref{7}  
\soulregister\cite{7} 
\renewcommand{\st}[1]{}

\usepackage{xr}

\makeatletter
\newcommand*{\addFileDependency}[1]{
  \typeout{(#1)}
  \@addtofilelist{#1}
  \IfFileExists{#1}{}{\typeout{No file #1.}}
}
\makeatother

\usepackage{textcomp} 
\usepackage{xifthen}
\usepackage{xcolor}
\usepackage{etoolbox}
\newboolean{togglechanges} 
\usepackage{siunitx}
\usepackage{wrapfig}

\setboolean{togglechanges}{false}

\newcommand{\comment}[1]{\ifbool{togglechanges}
    {#1}  
    {\textcolor{blue}{#1}}}

\usepackage{bibentry}

\usepackage{graphicx}
\usepackage{dcolumn}
\usepackage{bm}

\usepackage{textcomp} 

\begin{document}
\rmfamily

\title{Observing the dynamics of quantum states generated inside nonlinear optical cavities}
\author{Seou~Choi$^{\ddag,1}$}
\email{seouc130@mit.edu}
\author{Yannick~Salamin$^{\ddag,1,2,3}$}
\email{yannick.salamin@ucf.edu}
\author{Charles~Roques-Carmes$^{1,4}$}
\author{Jamison~Sloan$^{1,4}$}
\author{Michael~Horodynski$^{2}$}
\author{Marin~Solja\v{c}i\'{c}$^{1,2}$}
\affiliation{$^\ddag$ denotes equal contribution.\looseness=-1}
\affiliation{$^{1}$ Research Laboratory of Electronics, Massachusetts Institute of Technology, Cambridge, MA 02139, USA\looseness=-1}
\affiliation{$^{2}$ Department of Physics, Massachusetts Institute of Technology, Cambridge, MA 02139, USA\looseness=-1}
\affiliation{$^3$ CREOL, The College of Optics and Photonics, University of Central Florida, Orlando, Florida 32816, USA}
\affiliation{$^{4}$ E. L. Ginzton Laboratories, Stanford University, 348 Via Pueblo, Stanford, CA USA\looseness=-1}

\clearpage 

\setlength{\parskip}{0em}
\vspace*{-2em}


\vspace{0.8cm}

\begin{abstract}

      Quantum states at optical frequencies are often generated inside cavities to facilitate strong nonlinear interactions. However, measuring these quantum states with traditional homodyne techniques poses a challenge, as outcoupling from the cavity disturbs the state's quantum properties. Here, we propose a framework for reconstructing quantum states generated inside nonlinear optical cavities and observing their dynamics. Our approach directly imprints the field distribution of the cavity quantum state onto the statistics of bistable cavity steady-states. We propose a protocol to fully reconstruct the cavity quantum state, visualized in 2D phase-space, by measuring the changes in the steady-state statistics induced by a probe signal injected into the cavity under different condition. We experimentally demonstrate our approach in a degenerate optical parametric oscillator, generating and reconstructing the quasi-probability distribution of different quantum states. As a validation, we reconstruct the Husimi $Q$ function of the cavity squeezed vacuum state. In addition, we observe the evolution of the quantum vacuum state inside the cavity as it undergoes phase-sensitive amplification. By enabling generation and measurement of quantum states in a single nonlinear optical cavity, our method realizes an "end-to-end" approach to intracavity quantum tomography, facilitating studies of quantum optical phenomena in nonlinear driven-dissipative systems.    
\end{abstract}

\maketitle

\section*{Introduction} 

Quantum systems at optical frequencies are of significant interest in various applications including metrology~\cite{aasi2013enhanced}, computing~\cite{reimer2016generation}, and cryptography~\cite{zhao2006experimental}. A crucial aspect is the ability to generate and characterize quantum states such as single photon sources~\cite{michler2000quantum,claudon2010highly}, entangled photon pairs~\cite{wang2018multidimensional}, and squeezed states of light~\cite{nehra2022few}. Optical cavities play a crucial role by enabling enhanced control over light-matter coupling regimes~\cite{PhysRevLett.95.013904,frisk2019ultrastrong}, facilitating the creation of complex non-Gaussian states~\cite{PRXQuantum.2.030204,rivera2023creating}, and allowing the study of decoherence in quantum systems~\cite{deleglise2008reconstruction,kirchmair2013observation}.

Quantum tomography has become an essential tool to characterize quantum states~\cite{deleglise2008reconstruction, kalash2023wigner}. However, conventional tomography techniques based on homodyne detection~\cite{lvovsky2009continuous} -- which relies on interference of \textit{propagating} waves -- are not suitable for measuring quantum states confined within a cavity -- which are inherently \textit{stationary}. Furthermore, extracting the quantum states from a cavity typically alters their quantum properties, making accurate measurements difficult. 

To probe cavity quantum states, it is necessary to measure observables with sufficient sensitivity to detect subtle shifts induced by the cavity state, often involving just a few photons. For instance, neutral atoms have been used to probe cavity quantum states by detecting changes in the atom's energy state as they interact with cavity photons~\cite{deleglise2008reconstruction,gleyzes2007quantum,nogues1999seeing}. Alternatively, microwave cavity systems have taken advantage of the strong Kerr nonlinearity of a Josephson junction to induce measurable changes in coupling rate with a readout cavity, proportional to the photon number of the quantum state~\cite{kirchmair2013observation}. 

\begin{figure*}[t]
    \centering
    \includegraphics[scale = 1]{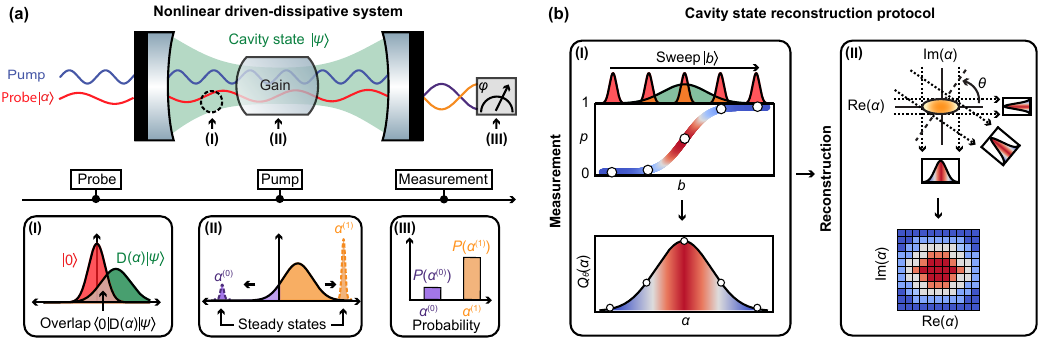}
    \vspace*{-4mm}
    \caption{\textbf{Measuring quantum states generated inside nonlinear optical cavities.} (\textbf{a}) Reconstructing cavity quantum states leveraging the sensitivity of a multistable system to weak coherent fields. The injected coherent field $\ket{\alpha}$ displaces the cavity state $\ket{\psi}$, changing the probability $p$ of measuring certain steady state. We measure the change in the probability $p$ to reconstruct the cavity state $\ket{\psi}$. (\textbf{b}) Visualizing cavity quantum states by measuring the Husimi $Q$ function. The sensitivity $\partial p/\partial b$ gives $Q_{\theta}$, which is the marginal distribution of the Husimi $Q$ function. $Q_{\theta}$ for different $\theta$ is measured to reconstruct the full 2D \textit{Q} function.}
    \label{fig:Fig1}
\end{figure*}

Achieving similar sensitivity at optical frequencies has proven to be very challenging due to inherently weak material nonlinearities. Nevertheless, nonlinear driven-dissipative systems such as optical parametric oscillators (OPOs) have revealed a wealth of intriguing non-classical behavior, including quantum states confined inside the cavity~\cite{park2024single} as well as out-coupled states of squeezed light \cite{aasi2013enhanced, yap2020generation}. In particular, studies have shown that macroscopic observables in similar multistable optical systems can be sensitive to fields at the single- and sub-photon levels~\cite{Arecchi1989,roques2023biasing, choi2024photonic}, opening new pathways for probing quantum optical systems. 

Here, we propose a framework for quantum tomography of optical quantum states generated within a nonlinear cavity featuring multiple steady states. Our method relies on mapping the initial field distribution of the cavity quantum state onto the statistics of the macroscopic cavity states. The mapping is achieved by measuring the system's sensitivity to an injected coherent field. Our approach enables the visualization of quasi-probability distribution of cavity quantum states in phase space representation. We demonstrate our concept in a degenerate optical parametric oscillator (DOPO) platform, and experimentally validate our method by reconstructing the Husimi $Q$ function of a squeezed vacuum state generated inside the cavity. We further showcase the potential of our method by directly observing the phase-sensitive amplification of the quantum vacuum state inside the cavity under high gain pumping condition. Our nonlinear driven-dissipative system supports both generation and measurement of cavity quantum states, facilitating ``end-to-end'' studies on intracavity phenomena such as non-classical state dynamics and quantum-to-classical transitions.

\section*{Results}
A common method for visualizing the quasi-probability distribution of a quantum state in 2D phase space is through the Husimi $Q$ function, expressed as ${Q(\alpha,\alpha^{\ast})=\frac{1}{\pi}|\braket{\alpha|\psi}|^2}$~\cite{carmichael2007statistical}. This function evaluates the overlap $\braket{\alpha|\psi}$ between a coherent state $\ket{\alpha}$ and the quantum state of interest $\ket{\psi}$ (see Supplementary Information Section~S1 for further details). 

Figure~\ref{fig:Fig1} illustrates the proposed method for measuring the overlap between a coherent state $\ket{\alpha}$ (red signal) and an unknown cavity quantum state $\ket{\psi}$ (green mode). The first step consists of coupling a coherent state into the cavity, in which the interaction with the quantum state leads to an increase in the mean electric field. The resulting state in the cavity can be described as a displacement $D(\alpha)$ of the cavity state $\ket{\psi}$ proportional to the injected coherent state $\ket{\alpha}$, as shown in panel I of Fig.~\ref{fig:Fig1}(a). 

To measure the overlap $\braket{\alpha|\psi}$, we leverage the sensitivity of a multistable system to its initial conditions. Specifically, we consider a DOPO pumped above threshold, in which two degenerate steady states $\alpha^{(0)}$ (purple signal) and $\alpha^{(1)}$ (orange signal) with phases $0$ and $\pi$ exist~\cite{marandi2012all}. Because the parametric amplification in DOPO is phase-sensitive, the initial cavity field distribution can be mapped on the statistics of the measured output steady states (panel II of Fig.~\ref{fig:Fig1}(a))~\cite{nehra2022few,aasi2013enhanced,roques2023biasing}. 
Hence, when the initial condition of the system is the quantum vacuum state $\ket{0}$, one observes an equal probability $p=p(\alpha^{(1)})$ and $1-p = p(\alpha^{(0)})$ of measuring either state (i.e., $p=0.5$)~\cite{marandi2012all}. 

On the other hand, when displacing the cavity state with a coherent state, the initial field distribution is displaced (e.g. shift to the right side in panel II of Fig.~\ref{fig:Fig1}(a)). This shift is directly translated in the measured state probability (e.g. $p> 0.5$ in panel III of Fig.~\ref{fig:Fig1}(a))~\cite{roques2023biasing}. Notably, the displacement also results in a change in the overlap between the quantum vacuum state and the cavity state, which can be described as $\braket{0|D(\alpha)|\psi}$ (panel I of Fig.~\ref{fig:Fig1}(a)). Therefore, we can analytically describe the probability $p$ as a function of the coherent field $b$ injected into the cavity as:   

\begin{equation}
\begin{aligned} \label{eq:overlap}
    p(b|\theta) = \int^{\infty}_{-D(b)} \frac{1}{\pi} |\braket{\alpha|\psi_\theta}|^2 d\alpha,
\end{aligned}
\end{equation}
where $\ket{\psi_\theta}$ is the 1D field distribution defined along the axis rotated by $\theta$ with respect to $\text{Re}(\alpha)$ on phase space. To avoid confusion between integration variable and the probe signal we inject to the cavity, we use $\ket{b}$ in Eq.~(\ref{eq:overlap}) to define the probe signal. For the rest of the discussion, we will refer to the injected probe field $b$ as a ``bias field'', as this field biases the probability distribution of measuring a certain steady state toward the value corresponding to the bias' phase. Derivation of Eq.~(\ref{eq:overlap}) can be found in the Methods section, and the dependence of phase-sensitive amplification behavior on the phase of the pump and the bias can be found in the Supplementary Information Section~S2.

\begin{figure*}[t]
    \centering
    \includegraphics[scale = 1.0]{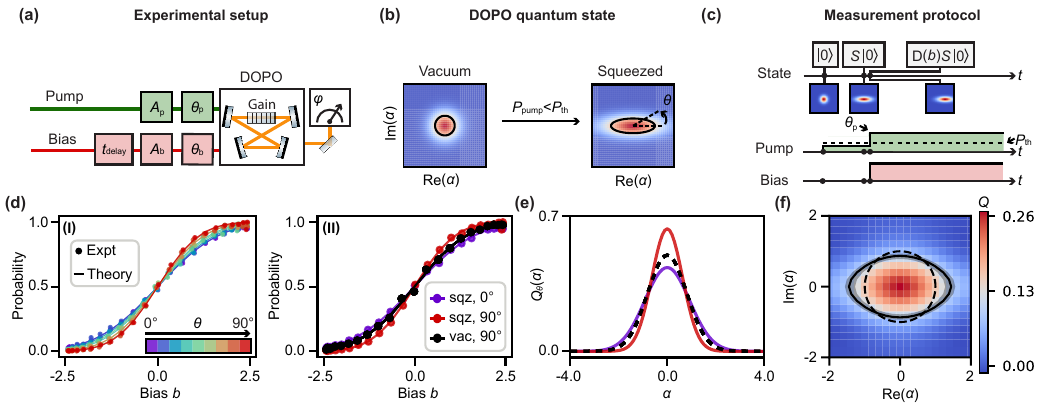}
    \vspace*{-4mm}
    \caption{\textbf{Measuring cavity squeezing.} (\textbf{a}) Schematic of the experimental setup. (\textbf{b}) Phase-sensitive amplification in nonlinear drive-dissipative systems. When the DOPO is pumped below threshold ($P_{\mathrm{pump}} < P_{\mathrm{th}}$), the noise for a certain quadrature can be reduced below the standard quantum limit. (\textbf{c}) Measurement protocol for a cavity squeezed vacuum state. Pump and bias are controlled to both generate and measure the cavity squeezed vacuum state. (\textbf{d}) Bias -- probability curves showing phase-sensitive amplification. (\textbf{e}) 
    Reconstructed marginal $Q$ function $Q_{\theta}$. Each curve is drawn by calculating the sensitivity from the bias -- probability curve. 
    (\textbf{f}) Reconstructed 2D $Q$ function. Solid (dashed) black line represents the contour line at $1/e$ for squeezed vacuum state (analytical solution of quantum vacuum state). Black shaded area corresponds to the measurement uncertainty in the contour line at $1/e$ for the squeezed vacuum state.}
    \label{fig:fig2}
\end{figure*}

Eq.~(\ref{eq:overlap}) shows that the probability $p$ is determined by the cumulative distribution of the cavity quantum state for $\alpha \geq -D(b)$ (or equivalently the orange area of the displaced quantum state in panel II of Fig.~\ref{fig:Fig1}(a)). It becomes evident that the sensitivity of the probability to the bias, $\partial p/\partial b$, can be used to reconstruct the marginal distribution of the Husimi $Q$ function at given $\theta$, which we denote as $Q_{\theta}$.

The protocol for reconstructing the Husimi $Q$ function of the cavity quantum state is depicted in Fig.~\ref{fig:Fig1}(b). The procedure starts from 
measuring the change in the probability $p$ by sweeping the bias field $b$ at given $\theta$. The sensitivity $\partial p/\partial b$ of the bias -- probability curve at a given $\theta$ provides the marginal distribution $Q_\theta$ (panel I of Fig.~\ref{fig:Fig1}(b)). This procedure is repeated to collect marginal distributions at different $\theta$ by rotating the pump phase (panel II of Fig.~\ref{fig:Fig1}(b)). Finally, an inverse Radon transform is applied to reconstruct the full 2D Husimi $Q$ function~\cite{deans2007radon}. See Supplementary Information Section~S2 for the detailed discussion of the reconstruction protocol. 


\subsection*{Measuring cavity squeezed vacuum state}

\begin{figure*}[t]
    \centering
    \includegraphics[scale = 1.0]{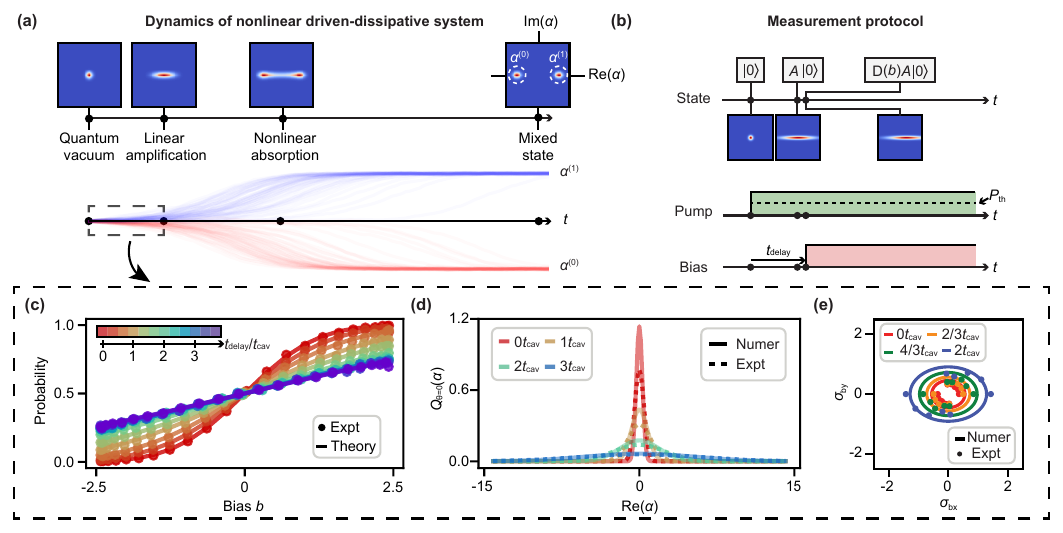}
    \vspace*{-4mm}
    \caption{\textbf{Visualizing the cavity dynamics of the DOPO.} (\textbf{a}) Dynamics of the quantum vacuum state in the DOPO. Stochastic trajectories (lower panel) and $Q$ function (upper panel) visualize the amplification and bifurcation of the cavity state. (\textbf{b}) Measurement protocol for an amplified quantum vacuum state. $A$ is an operator that maps the initial quantum vacuum state to its amplified state. (\textbf{c}) Bias -- probability curves at different bias injection time measured during the linear amplification stage. (\textbf{d}) Amplification of the quantum vacuum state described as broadening of the marginal \textit{Q} function $Q_{\theta=0}$. (\textbf{e}) Asymmetric amplification of the quantum vacuum state. $\sigma_{\text{b}}$ is the standard deviation of the sensitivity curve (i.e., $\partial p / \partial b$). Faster increase of $\sigma_{\text{b}}$ along the horizontal, which is aligned to the pump phase, shows that the quantum vacuum state amplifies along the pump phase.}
    \label{fig:Fig3} 
\end{figure*}

We first apply the quantum tomography protocol described in the previous section to measure cavity squeezing in a DOPO. A simplified schematic of our experimental setup is shown in Fig.~\ref{fig:fig2}(a), in which the DOPO consists of a nonlinear bow-tie cavity \cite{roques2023biasing}. In order to prepare and reconstruct different cavity quantum states, we implemented rapid control over both the amplitude $A_{\text{p}}$ and phase $\theta_{\text{p}}$ of the pump (green signal). To measure $\partial p / \partial b$ under different bias conditions, we have independent control of the amplitude $A_{\text{b}}$, phase $\theta_{\text{b}}$, and injection time $t_{\text{delay}}$ of the bias field $b$ (red signal). Details of the experimental setup can be found in the Methods section as well as in Supplementary Information Section~S4. 

As shown in Fig.~\ref{fig:fig2}(b), the noise of the initial quantum vacuum state undergoes phase dependent amplification. The pump is given along $\theta=0^{\circ}$ to amplify the quadrature noise along $\text{Re}(\alpha)$. For the quadrature orthogonal to the initial pump phase (i.e. $\theta=90^{\circ}$), de-amplification occurs, resulting in quadrature noise below the standard quantum limit~\cite{carmichael2013statistical}. We follow the protocol in Fig.~\ref{fig:fig2}(c) to generate and measure the cavity squeezed vacuum state. First, the DOPO is pumped below threshold $P_{\mathrm{th}}$ to create a squeezed vacuum state ${S}\ket{0}$. Next, the bias field is introduced to displace the squeezed vacuum state (${D}(b){S}\ket{0}$) and the pump power is simultaneously increased above threshold to amplify the displaced state. Steady-state outcomes are recorded to obtain the system's probability. This is repeated with varying the amplitude of the bias field to establish the bias -- probability relationship for a given pump phase. Finally, we measure the bias -- probability relationship at different $\theta$, by introducing an additional rotation of the pump phase when injecting the bias, allowing us to reconstruct the 2D Husimi $Q$ function.

The measured sensitivity of the probability to the bias at different $\theta$ is shown in the panel I of Fig.~\ref{fig:fig2}(d). As $\theta$ gets closer to 90$^{\circ}$, the probability becomes more sensitive to the bias field, showing that the phase-sensitive amplification happens inside the DOPO. In other words, a sharper transition is a result of a smaller cavity field amplitude along that quadrature. We also fitted the experimental data with the theoretically predicted bias -- probability curve which follows a Gauss error function, showing a good agreement with the data. Details of the theoretical model of the bias -- probability relationship can be found in the Methods section. The panel II of Fig.~\ref{fig:fig2}(d) highlights the amplification (purple curve) and deamplification (red curve) of the initial quantum vacuum state along different pump phase. Figure~\ref{fig:fig2}(e) shows the computed $Q_\theta$ functions for the two major quadratures. Comparing it to the $Q_\theta$ function of the quantum vacuum state, we observe squeezing and anti-squeezing of the $Q_\theta$ function at $\theta \sim 90^{\circ}$ and $\theta \sim 0^{\circ}$, respectively. 

Figure~\ref{fig:fig2}(f) shows the reconstructed 2D Husimi $Q$ function of the cavity squeezed vacuum state, with a squeezing ratio of $\sim 1.2 \pm 0.6$~dB along $\theta = 90^{\circ}$ obtained directly from the $Q_\theta$ functions. A squeezing of $\sim 1.8 \pm 0.7$~dB is estimated when accounting for experimental limitations such as the finite rise time of the pump signal, which pre-amplifies the squeezed vacuum state prior to the measurement. Detailed discussion on how including finite rise time of the pump signal allows us to get better estimation can be found in Supplementary Information Section~S5.  



To achieve maximum squeezing, which is known to be $3$~dB for cavity squeezed vacuum state~\cite{gardiner2004quantum}, the pump power must be just below the threshold during the quantum state preparation. In our current setup, we set the pump power at $\sim 80 \%$ of the threshold to provide a sufficient buffer and avoid amplification while preparing the squeezed vacuum state. Future improvement in using a low-noise laser and high bandwidth voltage signal, should bring us closer to the theoretical squeezing limit of $3$~dB. 

\subsection*{Observing intracavity dynamics}

Next, we use our reconstruction protocol to visualize the dynamics of the cavity quantum state. Figure~\ref{fig:Fig3}(a) illustrates how the initial quantum vacuum state evolves when the DOPO is pumped above the threshold. Stochastic trajectories show the time evolution of the DOPO signal amplitude, which are obtained by solving multiple times the SDEs derived from the original Hamiltonian (see Supplementary Information Section~S3 for further details). The cavity state undergoes two different stages before reaching steady-state -- a mixed state of two coherent states (see rightmost panel in Fig.~\ref{fig:Fig3}(a)). During the first stage, the parametric gain of the pump linearly amplifies the quantum vacuum state, broadening the distribution along one quadrature, i.e., Re($\alpha)$. As the photon number inside the cavity increases to the level where nonlinear absorption occurs, part of the signal field is converted back into the pump field. This marks the onset of the second stage, where the nonlinear absorption term becomes apparent, and the broadened Gaussian distribution bifurcates into two sidelobes. Once the nonlinear absorption and the linear amplification reach equilibrium, we observe a mixed state of coherent states with an equal probability. 

Using the reconstruction protocol, we can track the evolution of the Husimi $Q$ function, which manifests itself in the time dependence of the sensitivity of the probability to the bias field. Figure~\ref{fig:Fig3}(b) illustrates the measurement protocol to observe the amplification dynamics of the DOPO. The DOPO is pumped above threshold at $t=0$, amplifying the quantum vacuum state. After a time delay $t_{\text{delay}}$, the bias is introduced to measure the amplified state. By repeating the measurement for different delays, we can observe the cavity dynamics of the DOPO. In our experiment, we focus on the linear amplification stage (dashed box in Fig.~\ref{fig:Fig3}(a)) due to the finite extinction ratio of the bias amplitude modulator. Further discussions on how finite extinction ratio of the bias can affect the bias -- probability relationship can be found in the Supplementary Information Section~S5. 

Figure~\ref{fig:Fig3}(c) shows the measured bias-probability curves for different delays. It is evident that the system becomes less sensitive to the bias with increasing delay. Calculating the sensitivity of the bias -- probability curves, we can visualize the broadening of the marginal distribution of the Husimi $Q$ function along the pump phase $Q_{\theta=0}$ as Fig.~\ref{fig:Fig3}(d). As we expect from Fig.~\ref{fig:Fig3}(c), the amplified quantum vacuum state becomes less sensitive to the bias field as we need stronger bias field to fully displace it. We also measure the bias sensitivity of the amplified quantum vacuum at different pump phases (Fig.~\ref{fig:Fig3}(e)). We can clearly see the asymmetric phase-sensitive amplification along different pump phases. The additional amplification along the orthogonal axis is introduced because of the slow rise time of the voltage signal that controls the pump amplitude modulator. The measurement of bias -- probability curves at different pump phases and the detailed discussion on phase-sensitive amplification in Fig.~\ref{fig:Fig3}(e) can be found in Supplementary Information Section~S5.

\section{Discussion}

The presented framework leverages the phase-sensitive amplification in a nonlinear driven-dissipative system to generate a quantum state, and maps its field distribution onto the statistics of macroscopic cavity steady-states. This allows for an ``end-to-end'' cavity tomography experiment, supporting both generation and measurement of cavity quantum states. 

Our current experimental setup is limited to measuring quantum states that have a low photon number, such as squeezed vacuum state or linearly amplified quantum vacuum state. This limitation originates from two main factors: (1) the relatively low extinction ratio of the bias field; and (2) the slow rise time of the pump and bias signal due to the low bandwidth of the voltage amplifier that drives the electro-optic modulators. Further details on the origin of these experimental limitations are discussed in Supplementary Information Section~S4. We anticipate that integrated platforms could significantly improve both the response time and the extinction ratio~\cite{zhu2021integrated}. 

With a higher extinction ratio of the bias field, displacement of cavity states with a greater number of photons would become possible. This is crucial for observing the bifurcation process, where the distribution transitions from a Gaussian to Gaussian mixture (i.e., timeline between the second and third Husimi $Q$ functions in the upper panel of Fig.~\ref{fig:Fig3}(a)). 
Numerical experiments reconstructing the dynamics of the bifurcation is discussed in the Supplementary Information Section~S3. 

Improving the rise time of the optical signals would open the avenue for capturing the DOPO dynamics with a better time resolution. For instance, this would allow real-time observation of the generation and collapse of the squeezed vacuum state (see Supplementary Information Section~S3 for the corresponding numerical experiment). This would provide valuable insights into exotic quantum-optical phenomena in nonlinear driven-dissipative systems, including soliton generation~\cite{herr2012universal}, optical bistability~\cite{marandi2012all,roques2023biasing}, and Kerr frequency combs~\cite{zhou2015stability}. 

In conclusion, we have presented a framework for observing the dynamics of the quantum states generated inside optical cavities, experimentally realized in a DOPO system. We demonstrated a full reconstruction of a cavity squeezed vacuum state. In addition, we visualized the phase-sensitive amplification of the quantum vacuum state. Our method offers a robust approach for exploring cavity quantum phenomena in nonlinear driven-dissipative systems. 

\section{Methods}
\textbf{Experimental setup.}
Here, we briefly introduce our experimental setup. More detailed description of the setup can be found in Supplementary Information Section~S4.

\textit{Laser:} For the laser system, a Toptica FemtoFiber ultra 780 was customized to provide dual-wavelength femtosecond pulses, with \SI{780}{nm} and \SI{1560}{nm}. \SI{780}{nm} signal is used to pump the OPO, and \SI{1560}{nm} signal is split into two different lines, one is the bias field and the other is the local oscillator (LO) that is used to measure the phase of the OPO steady-state. Pump and bias field have the same polarization state, and they are orthogonal to the polarization of the LO signal. 


\textit{OPO:} The OPO consists of a MgO:PPLN crystal (MSHG1550-0.5-1, Covesion Ltd., United Kingdom) in a free-space bow tie optical cavity. The crystal is placed on the oven mount (PV10, Covesion Ltd., United Kingdom), which is connected to the temperature controller (OC2, Covesion Ltd., United Kingdom). The cavity length is matched to a single round trip of the laser pulse, and stabilized by detecting the power of the out-coupled sum-frequency mixed signal between the pump and the OPO signal~\cite{cheng2020dither}. The out-coupled sum-frequency mixed signal passes a narrow-band pass filter, which gives a linear change in the detected signal power as the cavity length deviates from the degenerate mode. A photodetector measures this error signal and a proportional-integral-derivative controller yields a voltage feedback to the piezoelectric actuator (PA44M3KW, Thorlabs, USA) attached to one of the cavity mirrors. To collect the statistics of the OPO steady-state and measure the probability, we modulate the pump and the bias signal at \SI{10}{kHz} using a lithium niobate electro-optic modulators (EO-AM-NR-C1, EO-AM-NR-C3, Thorlabs, USA for amplitude modulation, and EO-PM-NR-C1, EO-PM-NR-C3, Thorlabs, USA for phase modulation). The cavity lifetime is calculated from the loss of the OPO signal from each cavity mirror, which gives $\sim$\SI{300}{ns}.

\textit{Preparing bias signal:}
To match the temporal alignment between pump and bias pulses, we use an optical delay line with a linear translational stage. Additional precision could be achieved by using an electronically controlled motorized actuator (PIA25, Thorlabs, USA). To attenuate the bias signal before entering to the cavity, the bias field first passes a series of neutral density filters (NDC-100S-4, Thorlabs, USA). Then we use a combination of waveplates and polarizers to further attenuate the bias field.  The last waveplate is mounted on a motorized precision rotation stage (KPRM1E/M, Thorlabs, USA) which can be electronically controlled. 

\textit{Measuring bias field:}
Because the bias pulse entering the cavity contains less than a few photons, we use the following protocol to measure the amplitude of the bias field. First, we place the power meter after the last polarizer and set the waveplate angle that gives the maximum bias power $P_0$. Then the bias power at waveplate angle $\phi$ would be written as $P=P_0 \cos^{2}(2(\phi-\phi_0))$, where  $\phi_{0}$ is the angle that gives the maximum power. We rotate the angle of the waveplate to sweep the bias field and measure the bias -- probability curve. After passing the last polarizer, we calculate the actual bias power that interacts inside the crystal, considering all the possible power loss happening at each optical component before it reaches at the crystal. The loss of each element is obtained either from our previous measurements ~\cite{roques2023biasing} or the manufacturer. Then we calculate the bias field $b = \frac{E_\text{bias,mean}}{T_\text{rt}}\sqrt{\frac{\epsilon V}{\hbar\omega_{0}}}$. $E_\text{bias,mean}$ is the mean electric field of the bias, $T_\text{rt}$ is the round trip time of the cavity, $\epsilon$ is the electric permittivity of the MgO:PPLN crystal, $V$ is the mode volume, and $\omega_{0}$ is the angular frequency of the bias field. We use pulse duration of \SI{190}{ns}, beam waist of \SI{10}{\micro\metre} to calculate $E_\text{bias,mean}$ and $V$~\cite{roques2023biasing}. 

\textit{Measuring the statistics of the OPO steady-states:} The sampled OPO signal and the LO are combined in a standard interferometry setup to measure the phase of the OPO steady-state. The polarization states of the OPO signal and the LO are matched before they are combined with the beam splitter cube. We add an additional delay stage on the LO line, using the same module that we used to control the temporal alignment between the pump and the bias pulses. Using the one output arm of the beam splitter, we check the temporal alignment between the OPO signal and the LO by observing the inference pattern with the camera (CMLN-13S2M-CS, Edmund Optics, USA). For the other arm, we measure the interference pattern with the photodetector (PDA50B2, Thorlabs, USA). The detected signal was recorded on the oscilloscope, collecting either 1,000 (Fig.~\ref{fig:fig2}) or 10,000 OPO steady-states (Fig.~\ref{fig:Fig3}) to measure the probability at a given bias field. The probability is defined as the number steady-states with phase \SI{0}{rad} divided by the number of total steady-states measured with the oscilloscope~\cite{roques_zenodo}.   

\textbf{Reconstruction protocol.} In this section, we describe some mathematical background on our Husimi $Q$ function reconstruction protocol by measuring the sensitivity of the bias -- probability relationship.

\textit{Husimi Q function:} The Husimi $Q$ function gives the probability of observing a certain coherent state $\ket{\alpha}$ for a given quantum state $\ket{\psi}$:
\begin{equation}
\begin{aligned}\label{eq:Definition of Q representation}
Q(\alpha,\alpha^{\ast}) &= \frac{1}{\pi} \braket{\alpha|\rho|\alpha}\\
                        &= \frac{1}{\pi} |\braket{\alpha|\psi}|^2.
\end{aligned}
\end{equation}

Using a standard homodyne detection method, the Husimi $Q$ function can be measured by combining the unknown quantum state $\ket{\psi}$ with a reference coherent state $\ket{\alpha}$ using a beam splitter. Then the overlap can be measured using photon-number-resolving detectors~\cite{nehra2020generalized}. We rewrite Eq.~(\ref{eq:Definition of Q representation}) using the displacement operator $D(\alpha)$:
\begin{equation}
\begin{aligned} \label{eq:Q representation displaced}
   Q(\alpha,{\alpha}^{*}) &\equiv \frac{1}{\pi}|\braket{\alpha|\psi}|^2\\
                          &= \frac{1}{\pi}|\braket{0|D(-\alpha)|\psi}|^2.
\end{aligned}
\end{equation}
Eq.~(\ref{eq:Q representation displaced}) shows that measuring the Husimi $Q$ function is equivalent to measuring the overlap between the displaced quantum state $D(-\alpha)\ket{\psi}$ and quantum vacuum state $\ket{0}$. 

\textit{Displacement operator:} The stochastic differential equation of the OPO signal defined along the real quadrature of the phase space $X$ can be written as below~\cite{roques2023biasing}:

\begin{equation}
\begin{aligned}\label{eq:1D SDE_X}
\dot{X} = (\lambda-1)X+\sqrt{2}b+\sqrt{\lambda}\eta_{X}.
\end{aligned}    
\end{equation}

Here, $\eta_{X}$ is the Gaussian noise. Because the derivative of the drift term $(\lambda-1)X + \sqrt{2}b$ with respect to $X$ is positive when the OPO is pumped above threshold (i.e. $\lambda > 1$), the sign of the drift term at the time when the bias field is injected determines whether $X$ will grow exponentially toward positive or negative side. Therefore, the direction of the amplification can be determined by the sign of the drift term. The critical point $X_{0}$ with zero drift happens at $X_{0}=-\sqrt{2}b/(\lambda-1)$. $X_{0}=0$ for an unbiased OPO, meaning that the amplification behavior of the cavity quantum state $\ket{\psi}$ in the biased OPO is identical to that of displaced quantum state $D(b)\ket{\psi}$ in the unbiased OPO, where the displacement operator $D(b)$ defined along the $X$ quadrature is written as:
\begin{equation}
\begin{aligned}\label{eq:displacement operator}
D(b) \equiv \frac{\sqrt{2}b}{\lambda-1}.
\end{aligned}    
\end{equation}

\textit{Reconstructing marginal distribution of the Husimi $Q$ function:} After the bias field displaces the quantum state, the OPO signal is amplified until it reaches one of the steady-states. As we described in the previous section, the original quantum state $\ket{\psi}$ in the biased OPO shows the same dynamics as $D(b)\ket{\psi}$ in the unbiased OPO system. The parametric amplification process is phase-sensitive, so $X >0$ (or equivalently $\text{Re}({\alpha})>0$)  part of the displaced quantum state gives the probability $p$ of measuring a certain steady-state, which can be analytically described as:  

\begin{equation}
\begin{aligned} \label{eq:ab initio probability}
   &p(b|\theta) \equiv \int^{\alpha=\infty}_{\alpha=0} { |{{\psi_{\theta,D(b)}}(\alpha)}|^2}d \alpha,
\end{aligned}
\end{equation}

where $\ket{\psi_{\theta,D(b)}} \equiv D(b)\ket{\psi_{\theta}}$. Because the phase-sensitive amplification happens along one quadrature, we consider the marginal distribution $\ket{\psi_{\theta}}$. Using basic algebra, 
\begin{equation}
\begin{aligned} \label{eq:probability to Q}
   p(b|\theta) & \equiv \int^{\infty}_{-D(b)}  { |{\psi_{\theta}(\alpha)}|^2}d\alpha \\
        & = \braket{\psi_{\theta}|\psi_{\theta}}_{[-D(b),\infty]}\\
        & = \braket{\psi_{\theta}|I|\psi_{\theta}}_{[-D(b),\infty]} \\
        & = \frac{1}{\pi} \int|\braket{\psi_{\theta}|{\alpha}}|_{[-D(b),\infty]}^{2} ~ d^2\alpha\\
        & = \int^{\infty}_{-D(b)} \int^{\infty}_{-\infty} {Q}~\text{dIm} (\alpha_{\theta})\text{dRe}{(\alpha_{\theta})}. \\
\end{aligned}
\end{equation}

We introduced the identity operator $I=\frac{1}{\pi} \int \ket{\alpha}\bra{\alpha}\text{d}^2\alpha$, to expand the representation from a specific coordinate to the 2D phase space. The final expression represents the same probability but now sums over the possible coherent states in the region of interest. Therefore, by taking the derivative with respect to $D(b)$, we can reconstruct the marginal distribution $Q_{\theta}$.

\textit{Stochastic differential equations:} Until now, we have focused on using the displacement operator to understand how the bias field $b$ affects the probability $p$. In this section, we provide another approach, which solves the SDE of the OPO system to calculate the bias -- probability relationship. When a time-dependent bias field $b(\tau)$ is introduced, the OPO dynamics at the linear amplification stage can be written as:
\begin{align}\label{eq:1D biased OPO SDE}
     \dot{X} = (\lambda-1)X + \sqrt{2}b(\tau)+\sqrt{\lambda}\eta_{X}.
\end{align}
With $Z=Xe^{-(\lambda-1){\tau}}$,
\begin{align}\label{eq:1D integral form}
     Z(\tau) = Z(0) + \int^{\tau}_{0}{e^{-(\lambda-1)\tau\prime}[\sqrt{2}b(\tau\prime)+\sqrt{\lambda}\eta_{X}(\tau\prime)]d\tau\prime}.
\end{align}
Calculating the positive area of the probability distribution of $Z(\tau)$, we can calculate the probability $p$. When a constant bias field $b_{0}$ is injected at $\tau=\tau_{0}$, 

\begin{align}\label{eq:1D integral of f}
     P[Z(\tau) \geq 0] = P\left[f(\tau)\geq -\frac{\sqrt{2}b_{0}}{\sqrt{\lambda}}\frac{e^{-(\lambda-1)\tau_{0}}-e^{-(\lambda-1)\tau}}{\lambda-1}\right]_{\textstyle ,}
\end{align}

\begin{align}\label{eq:definition of f}
 f(\tau)\equiv\int^{\tau}_{0}{e^{-(\lambda-1)\tau\prime}}\eta_{X}(\tau\prime)d\tau\prime 
 \end{align}

Because $X(\tau)$ is the Wiener process (linear superposition of Gaussian noises), $f(\tau)$ is a Gaussian with mean $\langle f(\tau) \rangle =0$, and variance $\langle f^{2}(\tau)\rangle=\frac{1-e^{-2(\lambda-1)\tau}}{2(\lambda-1)}$. Therefore, 
\begin{equation}\label{eq:1D probability becomes gaussian integral}
\begin{aligned}
    P[Z(\tau)\geq0]=\int^{\infty}_{-f_{0}(\tau)}{\frac{1}{\sqrt{2\pi \langle f^{2}(\tau)\rangle }}e^{-\frac{f^{2}}{{2\langle f^{2}(\tau)\rangle}}}}df
\end{aligned}_{\textstyle ,}
\end{equation}
where $f_{0}(\tau) \equiv \frac{\sqrt{2}b_{0}}{\sqrt{\lambda}}\frac{e^{-(\lambda-1)\tau_{0}}-e^{-(\lambda-1)\tau}}{\lambda-1}$. As we measure the probability $p$ after several cavity cycles, which corresponds to the time $\tau$ when $e^{-(\lambda-1)\tau} \rightarrow 0^{+}$, so the probability $p$ becomes 
\begin{equation}
\begin{aligned}\label{eq:1D biased p(b) at early stage}
    p \equiv P[Z(\tau)\geq0]=\frac{1}{2}\left( 1+\text{erf}\left[ \frac{\sqrt{2}b_{0}e^{-(\lambda-1)\tau_{0}}}{\sqrt{\lambda}\sqrt{\lambda-1}} \right] \right)_{\textstyle ,}
\end{aligned}
\end{equation}

where $\text{erf}$ is the Gauss error function. In the Supplementary Information Section~S3 and~S5, we discuss discuss the SDEs for time-dependent $\lambda(\tau)$ and $b(\tau)$.

\section{Authors contributions}
S.~C., Y.~S., C.~R.-C., J.~S., and M.~S. conceived the original idea. S.~C. and Y.~S. built the experimental setup with contributions from C.~R.-C., M.~H.;~S.~C. and Y.~S. acquired and analyzed the data. S.~C. developed the theoretical and numerical tools, with contributions from Y.~S., C.~R.-C., J.~S. and M.~H. The manuscript was written by S.~C., Y.~S., and C.~R.-C., with inputs from all authors.

\section{Competing interests}
The authors declare no potential competing financial interests.

\section{Acknowledgements}
The authors thank Alex Gu for stimulating discussion on phase space representation of cavity quantum states. S.~C. acknowledges support from Korea Foundation for Advanced Studies Overseas PhD Scholarship. Y.~S. acknowledges support from the Swiss National Science Foundation (SNSF) through the Early Postdoc Mobility Fellowship No.~P2EZP2-188091. C.~R.-C. is supported by a Stanford Science Fellowship. M.~H. acknowledges funding by the Austrian Science Fund (FWF) through grant J4729. J.~S. acknowledges earlier support from a Mathworks Fellowship. The authors acknowledge the MIT SuperCloud and Lincoln Laboratory Supercomputing Center for providing computation resources supporting the results reported within this paper. This material is based upon work also supported in part by the U. S. Army Research Office through the Institute for Soldier Nanotechnologies at MIT, under Collaborative Agreement Number W911NF-23-2-0121. This publication was also supported in part by the DARPA Agreement No. HO0011249049.

\end{document}